\documentstyle[12pt]{article}
\begin{document}
\begin{center}
{\Large\bf SOLITON SOLUTIONS OF THE $\sigma$ MODEL AND THE 
DISORIENTED CHIRAL CONDENSATES}
\vskip 0.2cm
Prasanta.~K.~Panigrahi\footnote{ email: panisp@uohyd.ernet.in} 
and C.~Nagaraja Kumar 
\vskip0.2cm
School of Physics, University of Hyderabad, Hyderabad - 500 046 
(INDIA).
\end{center}
\vskip 0.2cm
\begin{center} 
{\bf ABSTRACT}
\end{center}
We derive travelling soliton solutions of the $\sigma$ model,
relevant for the description of dynamics of pions, in the
high-energy, heavy-ion collisions. These extended objects are
analogs of Ne$\grave{e}$l and Bloch type defects in
ferromagnetic systems and could possibly describe the
disoriented chiral condensates.  It is shown that these
solutions are metastable and can naturally produce an asymmetry
between $\pi^0 $ and $ \pi^+$, $\pi^- $ mesons in the soft-pion
emissions from heavy-ion collisons.

The formation of the disoriented chiral condensates (DCC) in the
high-energy, heavy-ion collisions and their detection through
the asymmetry in the soft $\pi^0$ and $\pi^+ , \pi^-$ emissions
have generated considerable interest in the recent literature
[1,2,3].  There are also experimental indications pointing to
the possible formation of DCC in cosmic ray collisions[4].
Broadly speaking, DCC refers to the domains in space-time, where
the chiral order parameter points along a direction in the
iso-spin space, different from the iso-spin singlet
$\sigma$-direction in the outside physical vacuum. It is by now
well accepted that the $\sigma$-model with four fields $\sigma,
\pi^+, \pi^- $ and $\pi^0 $ having an approximate SU(2)$\otimes$
SU(2) symmetry effectively captures the low-energy dynamics of
QCD with two flavors. Below a certain critical temparature
$T_c$, the global SU(2) chiral symmetry is spontaneously broken
due to the non-vanishing vacuum expectation value (VEV) of the
iso-singlet fermion bilinear $< \bar\psi \psi >$.  In the
$\sigma$ model description, this corresponds to the $\sigma$
field having a VEV and further more the pions are realized as
the Goldstone modes in the broken symmetry phase.  The chiral
symmetry of the QCD Lagrangian is only approximate because of
the presence of the current quark masses and hence the pions
acquire a small but non-vanishing mass.

In high energy, heavy-ion collisions, there exists a possibility
that  chiral symmetry gets restored in a small region of hot
hadronic matter in the central  zone.  This region surrounded by
an outwardly expanding hot hadronic shell cools rapidly to a
temperature below $T_c$; the chiral order parameter in this
domain can in priniciple point in an isospin direction different
from the $\sigma$ field.  The eventual  realignment with the
physical vacuum at a later period in time can result in the
emission of soft, coherent pions.  There could be anomalous
fluctuations in neutral to charged pion ratio.  If the DCC
domain is not baryon rich the dynamics can be described by  the
$\sigma$ model.  Both the linear and the non-linear $\sigma$
models have been extensively used to study the formation of DCC
near $T_c$.

In the following, we follow the non-equilibrium phase transition
picture of DCC formation originally due to Rajagopal and Wilczek
[5].  Here quenching results in the removal of $\sigma$ and
$\vec \pi$ fields from contact with the heat bath. These fields
then evolve according to the $ T = 0 $ Hamiltonian.  It has been
shown numerically, in the context of the linear $\sigma$ model
that in intermediate times, $5 \sim 50 \times {1 \over m_\pi} $,
there is a dramatic amplification of long wavelength modes.
Since short wavelength modes are not completly absent, the
resulting domains are not smooth [6].

Hence the emerging   physical picture of DCC  is more closer to
a `pion laser' where the order parameter instead of being a
constant throughout the DCC can fluctuate and hence is
space-time dependent. The nonlinearity of the $\sigma$ model
makes it difficult to get the analytic time evolution in (3 + 1
) dimensions.  Furthermore mean field approximation may not be
fully reliable for the entire intermediate  time range. Keeping
these in mind, we start with  the idealized Heisenberg type
boundary condition : the thin disc representing the
Lorentz-contracted nuclei at the time of collision is infinite
in extent in the transverse direction .  Assuming the fields to
be independent of the transverse directions one can simplify the
problem to a ( 1 + 1 ) dimensional field theory [7] .  Inspired
by effective field theories describing lasers [8] and other
non-equilibrium phenomena [9] we consider the $\sigma$ model
Lagrangian in the presence of additional isospin violating
quadratic terms in the field variables and work out the
consequences.  These terms could possibly arise due to
non-equilibrium effects or squeezing [10]. The relevant
Lagrangian then is
\begin{equation}
{\cal L} ={{f_\pi}^2\over 2} [(\partial_\mu \sigma)^2 +
(\partial_\mu \pi^{a})^2 ] - V ( \pi^a , \sigma )
\end{equation}
\begin{equation}
V ( \pi^a , \sigma ) = \lambda \{ \sigma^2 + {\pi^a}^2 - 1 \}^2
- {\mu\over 2}  {\pi_3}^2 + \kappa
\end{equation}
$\kappa $ was added to adjust the lowest value of the potential
and the term with coefficient $\mu$ is the new term. The
$\pi_\pm $ are ${1\over \sqrt{2}} ({ \pi_1 \pm i \pi_2})$.  It
is easy to solve for the solutions for the equations of motion
and we find
\begin{eqnarray}
\pi_1& =l  ~~sech {\sqrt{2 \mu }\over f_\pi} \xi \\
\pi_2& = m ~~sech{\sqrt{2 \mu }\over f_\pi} \xi \\
\pi_3&= \alpha  \tanh{\sqrt{2 \mu }\over f_\pi} \xi \\
\sigma & = n ~~sech{\sqrt{2 \mu }\over f_\pi} \xi 
\end{eqnarray}
\begin{equation}
{\rm where} \qquad \qquad \alpha^2 = 1 + {\mu \over { 2
\lambda}}
\end{equation}
\begin{equation}
{\rm and}\qquad \qquad  l^2 + m^2 + n^2 = 1 - { \mu \over{ 2
\lambda}}
\end{equation}
this last relation puts a restriction on the values $\mu $ can
take to be less than $ 2 \lambda$.  The integrated energy
density for large distance $L$ is given by
\begin{equation}
\sqrt{\mu\over 2} f_\pi [ 2 - {\mu\over{ 3 \lambda}}]
\end{equation}
The ratio of the number of neutral pions to the total pions
produced  is proportional to   ratio of the squared amplitudes
of the corresponding pion fields integrated over the length L.

\begin{equation}
{N({\pi_0})\over N_{\rm total}} = {{{\int {\pi_3}^2} dx} \over
{\int ({\pi_a}^2 +
\sigma^2) dx }} = {\alpha^2 (  L - {Tanh(aL) \over a})  \over ({\alpha^2 L  - 
{\mu\over \lambda} {Tanh(a L )\over a}})}
\end{equation}
where $a$ is $ {\sqrt{2 \mu }\over f_\pi}$. For  small L this
ratio tends to zero and for large distances it gives the value
1.

Another  interesting term which can be added is $ \mu{ \pi_1}^2
= \mu {1\over 2} ( \pi_+ +  \pi_ -)^2 $. This term leads to
Bloch type solitons given by $ A Tanh (m\xi) + i B Sech (m\xi) $
which  smoothly connects $\pi_+ $ with $\pi_-$ asymptotically.
However, this term violates $U(1) $ electromagnetic  invariance
and hence may not be relevant in the present situation.

In conclusion, the soliton solutions which appear in other
fields of non-equilibrium pheonmena can possibly play a
significant role in the DCC formation.  A number of
investigations e.g., stability analysis, coupling with fermions
needs to be carried out for illustrating their role [11].  A
first principle study of the origin of the additional terms
starting from non-equilibrium field theory also needs to be
carried out.\\ {\bf Acknowledgments}: One of us (P.K.P) would
like to thank Drs. R. MacKenzie and M. Paranjape for hospitality
and Drs. K. Rajagopal and  T.Cohen for many discussions and
clarifications. (C.N.K) research is supported by CSIR, India,
through S.R.A. Scheme.

\noindent{\bf References}
\begin{enumerate}
\item [1] F. Wilczek, Int. Journ. Mod. Phys. {\bf A7 }, 3911
   (1992), J.Bjorken, Int. J. Mod. PHys. {\bf A7}, 4189 (1992)
\item [2] K.Rajagopal and F. Wilczek, Nucl. Phys. {\bf B399}, 395 (1993).
\item[3] A.Anslem, Phys. Lett. {\bf B 217 } 169, (1988).
\item [4] Chacaltaya-Pamir Collaboration; Tokyo University
preprint ICRR -Report -258-91-27 and references there in
\item[5] K. Rajagopal, preprint,  hep-ph  9703258, 
J.D. Bjorken, K.L.Kowalski and C.C.Taylor SLAC Preprint SLAC
-Pub-6109.
\item[6] G. Amelino-Camelia, J.D. Bjorken and S.E. Larsson, preprint
SLAC-PUB-7565, hep-ph 9706530.
\item[7] J.-P. Blaizot and A. Krzywicki, 
Phys. Rev. {\bf D 46} 246, (1992)
\item[8] S.Longhi, Europhysics Letters {\bf 37}, 257, (1997).
\item[9] P. Coullet and K. Emilsson Physica {\bf D 61} 119, (1992).
\item[10] B.A. Bambah, preprint, hep-ph 9708414 and references therein.
\item[11] P.K.Panigrahi and C.N.Kumar, manuscript under preparation.
\end{enumerate}
\end{document}